\documentclass{aip-cp}

\usepackage[numbers]{natbib}
\usepackage{rotating}
\usepackage{graphicx}
\usepackage{url}

\newcommand{\pks}{PKS~1510-089\ }

\begin{document}

\title{The Complex VHE And Multiwavelength Flaring Activity Of The FSRQ PKS 1510-089 In May 2015}

\author[aff1,aff2]{M.~Zacharias\corref{cor1}}
\author[aff2]{M.~B\"ottcher} 
\author[aff3]{N.~Chakraborty}
\author[aff1]{G.~Cologna}
\author[aff1]{F.~Jankowsky}
\author[aff4]{J.-P.~Lenain}
\author[aff1]{M.~Mohamed}
\author[aff5]{H.~Prokoph}
\author[aff1]{S.~Wagner}
\author[aff6]{A.~Wierzcholska}
\author[]{D.~Zaborov$^7$, for the H.E.S.S. Collaboration}

\affil[aff1]{Landessternwarte, Universit\"at Heidelberg, K\"onigstuhl, D 69117 Heidelberg, Germany}
\affil[aff2]{Centre for Space Research, North-West University, Potchefstroom 2520, South Africa}
\affil[aff3]{Alexander von Humboldt Fellow at Max-Planck-Institut f\"ur Kernphysik, P.O. Box 103980, D 69029 Heidelberg, Germany}
\affil[aff4]{Sorbonne Universit\'es, UPMC Universit\'e Paris 06, Universit\'e Paris Diderot, Sorbonne Paris Cit\'e, CNRS, Laboratoire de Physique Nucl\'eaire et de Hautes Energies (LPNHE), 4 place Jussieu, F-75252, Paris Cedex 5, France}
\affil[aff5]{Department of Physics and Electrical Engineering, Linnaeus University,  351 95 V\"axj\"o, Sweden}
\affil[aff6]{Instytut Fizyki J\c{a}drowej PAN, ul. Radzikowskiego 152, 31-342 Krak{\'o}w, Poland}
\affil[aff7]{Laboratoire Leprince-Ringuet, Ecole Polytechnique, CNRS/IN2P3, F-91128 Palaiseau, France}
\corresp[cor1]{E-mail: m.zacharias@lsw.uni-heidelberg.de}

\maketitle

\begin{abstract}
The blazar PKS 1510-089 was the first of the flat spectrum radio quasar type, which had been detected simultaneously by a ground based Cherenkov telescope (H.E.S.S.) and the LAT instrument on board the Fermi satellite. Given the strong broad line region emission defining this blazar class, and the resulting high optical depth for VHE ($E>100\,$GeV) $\gamma$-rays, it was surprising to detect VHE emission from such an object. In May 2015, PKS 1510-089 exhibited high states throughout the electromagnetic spectrum. Target of Opportunity observations with the H.E.S.S. experiment revealed strong and unprecedented variability of this source. Comparison with the lightcurves obtained with the \textit{Fermi}-LAT in HE $\gamma$-rays ($100\,$MeV$<E<100\,$GeV) and ATOM in the optical band shows a complex relationship between these energy bands. This points to a complex structure of the emission region, since the one-zone model has difficulties to reproduce the source behavior even when taking into account absorption by ambient soft photon fields. It will be shown that the presented results have important consequences for the explanation of FSRQ spectra and lightcurves, since the emission region cannot be located deep inside the broad line region as is typically assumed. Additionally, acceleration and cooling processes must be strongly time-dependent in order to account for the observed variability patterns.
\end{abstract}

\section{INTRODUCTION}
Flat spectrum radio quasars (FSRQs) are blazars with broad emission lines in the optical spectrum. This means that the nuclear region surrounding the central black hole is filled with hot and rapidly moving material emitting these Doppler broadened lines, the so-called broad line region (BLR). Any very high energy $\gamma$-ray emission (VHE, $E>100\,$GeV) emitted by the jet within the BLR is absorbed by the large number of optical photons. Thus, for many authors FSRQs, like 3C~279 and PKS~1510-089, were not expected to be detectable at VHE by ground-based Cherenkov telescopes. However, both mentioned sources were the first two FSRQs to be detected at VHE \cite{aeaM08,Hea13}, and at the time of writing six FSRQs are known to be VHE emitters.\footnote{The H.E.S.S. discovery of the sixth FSRQ, PKS~0736$+$017, has been announced at this conference; see \cite{cea16}.}

PKS~1510-089 (redshift $z=0.361$) is known for its complex multiwavelength behavior exhibiting both correlated and uncorrelated, i.e. orphan, flaring events in different energy bands. This is difficult to understand in the standard one-zone paradigm, where a single compact region is responsible for most of the output of the jet. Modeling the spectral energy distribution (SED) in a quiescent state has been shown to be difficult within the one-zone scenario leading several authors to believe that the one-zone model cannot hold for PKS~1510-089 \cite[e.g.,][]{nea12,b13}. Additionally, the detection of VHE emission by the H.E.S.S. experiment \cite{Hea13} indicates that at least one $\gamma$-ray emission region must be located at the edge or beyond the BLR in order to avoid the strong attenuation by the BLR photons \cite[e.g.,][]{bea14}. The VHE photons are thus expected to be inverse Compton scattered photons of the dusty torus.

From March to August 2015 \pks was very active in all energy bands. This was particularly evident in the optical R-band where monitoring with ATOM, the $75\,$cm class optical support instrument of H.E.S.S., revealed an unrivaled, long-lasting high state. \pks exhibited high optical fluxes for several weeks, which have only exceeded once in all previous ATOM observations since 2007. A contemporaneous high state in $\gamma$-rays observed by the \textit{Fermi}-LAT resulted in Target-of-Opportunity observations with the H.E.S.S. experiment in May 2015. 
Here the results of this campaign are presented, which show a very active source in VHE, exhibiting night-by-night variability. This behavior has not been detected before at VHE from PKS~1510-089. This result, along with multiwavelength data from \textit{Fermi}-LAT, \textit{Swift}, and ATOM, could be difficult to reproduce with the one-zone model and may favor a multi-zone scenario. 

\section{OBSERVATIONS}
\subsection{H.E.S.S.}
The High Energy Stereoscopic System (H.E.S.S.) is an array of five imaging air Cherenkov telescopes located in the Khomas Highland in Namibia. The original array consisting of four $12\,$m telescopes arranged in a square of $120\,$m side length has been upgraded in 2012 by the addition of a $28\,$m telescope in the center of the original square. The telescopes are optimized to detect the brief Cherenkov flashes generated when incoming VHE $\gamma$-rays interact with the Earth's atmosphere.

H.E.S.S. observed \pks in seven nights of May 2015 with the full array resulting in a detection significance of $18.2\,\sigma$ in $15.3\,$hrs of good quality data. Only stereo events were selected for the presented results implying that at least two telescopes had to register a Cherenkov shower. The analysis was conducted using the \textit{Model} analysis chain \cite{nr09}. An analysis with the independent analysis chain \textit{ImPACT} \cite{ph14} confirmed the results.

\subsection{Multiwavelength data}
In addition to the H.E.S.S. observations data was gathered in the $\gamma$-ray and X-ray bands, from the \textit{Fermi} and \textit{Swift} satellites, which are publicly available. Standard analysis techniques were used to derive the spectra and lightcurves. Optical data in the R-band were obtained with ATOM.

\section{MULTIWAVELENGTH LIGHTCURVE PROPERTIES}
\begin{figure}
\centering{\includegraphics[width=0.6\textwidth]{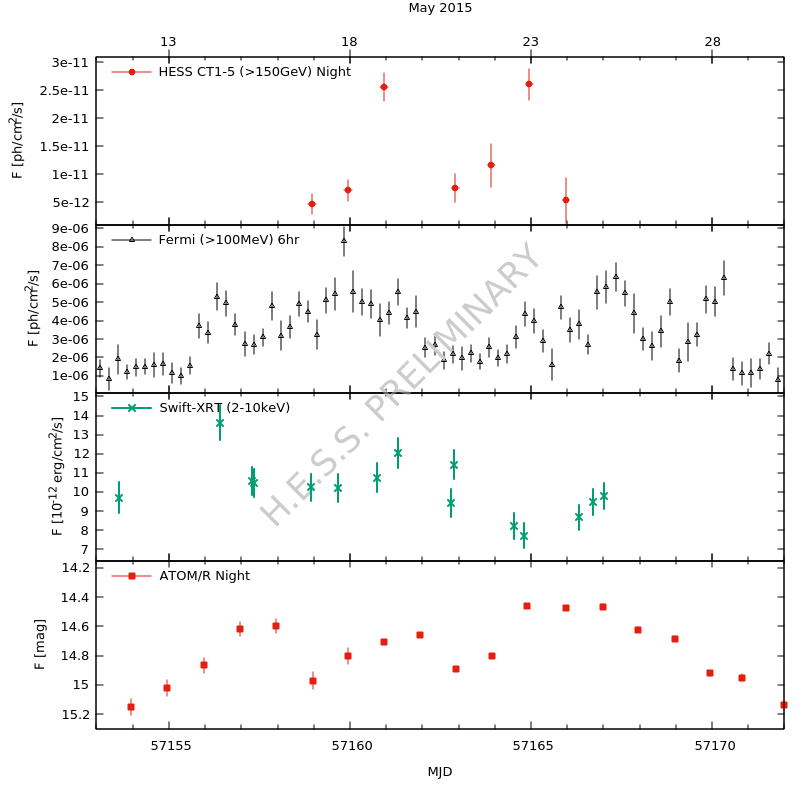}}
\caption{Multiwavelength lightcurves around the H.E.S.S. ToO campaign. 
\textit{Top:} VHE lightcurve from H.E.S.S. observations in nightly binning above an energy of $E>150\,$GeV.
\textit{2nd:} Lightcurve from \textit{Fermi}-LAT observations in $6\,$hr binning above an energy of $E>100\,$MeV.
\textit{3rd:} Lightcurve from \textit{Swift}-XRT observations for each pointed observation in the energy range $2\,$keV$<E<10\,$keV.
\textit{Bottom:} Lightcurve from ATOM R-band observation in nightly averages.}
\label{fig:lc}
\end{figure} 
The multiwavelength lightcurves of \pks during a time frame of about 2 weeks in May 2015 are shown in Fig. \ref{fig:lc}. The source is clearly variable in all bands.

The H.E.S.S. lightcurve shows variability on a night-by-night timescale. A constant flux is disfavored by more than $9\sigma$. This is the first time that night-by-night variability is reported for this source in the VHE band. The flux varies by a factor of $5$ over the observation period. The flux doubling time scale, defined as \cite{zea99}
\begin{eqnarray}
 t_{var} = \frac{F_1+F_2}{2} \frac{t_2-t_1}{|F_2-F_1|} \label{eq:tvar},
\end{eqnarray}
can be found to as short as $t_{var} = 18\,$hrs. 

The lightcurve as observed with the \textit{Fermi}-LAT also shows strong variability, even though the maximum flux is less than half the value of the historical maximum from 2012. The source varies by a factor $8$ over the shown observation period. Interestingly, there is no obvious correlation between the \textit{Fermi}-LAT lightcurve and the VHE lightcurve, even if one considers the different integration times of the flux points. The flux doubling time can be found as short as $t_{var} = 5\,$hrs. The photon index of the shown period is on average $\Gamma=2.21\pm 0.01$. The spectrum is, thus, harder than the typical spectrum with a photon index of $\Gamma = 2.36\pm 0.01$. This could be interpreted as a shift of the peak energy or as higher amplitude variability at higher energies.

The X-ray flux as measured by \textit{Swift}-XRT is also variable over the shown time range. The flux varies within a factor of 2. Despite the different sampling rates of the VHE and the X-ray band, there is also no obvious correlation between these two bands.

The R-band optical lightcurve from ATOM observations is shown as nightly averages, and varies by almost a magnitude in the entire period. Even more remarkable, the low-state R-band flux is typically below $16\,$mag. Before the 2015 observing season the R-band flux observed by ATOM has exceeded $15\,$mag. only once and for less than $10\,$days. Hence, this extended high state in the synchrotron component is very unusual. The optical observations were timed to match the observation times at VHE. While the maxima in the R-band are broader than in the VHE band, the initial increase in flux roughly coincide.

The correlation statements require more detailed analyses than possible in these proceedings, but they already point to a complex structure in the emission region.

\section{SED AND ABSORPTION FEATURES}
\begin{figure}
\begin{minipage}[t]{0.48\textwidth}
\centering \resizebox{\hsize}{!}
{\includegraphics{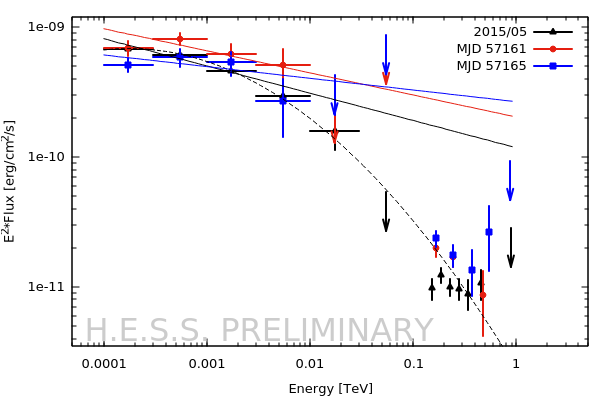}}
\end{minipage}
\hspace{\fill}
\begin{minipage}[t]{0.48\textwidth}
\centering \resizebox{\hsize}{!}
{\includegraphics{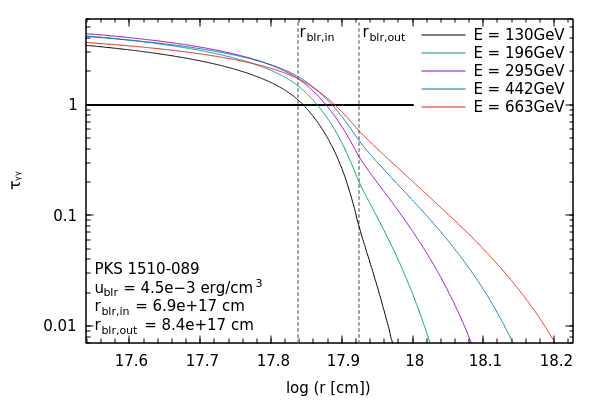}}
\end{minipage} 
\caption{\textit{Left:} H.E.S.S. and \textit{Fermi} EBL-deabsorbed spectra for three time intervals (\textit{black:} MJD 57155.5 - 57170.5; \textit{red:} MJD 57160.4 - 57161.4; \textit{blue:} MJD 57164.4 - 57165.4). The solid lines show an unabsorbed power-law extrapolation of the \textit{Fermi} data to the VHE range, while the dashed line is an example of the unabsorbed log-parabola extrapolation of the \textit{Fermi} data to the VHE.
\textit{Right:} Optical depth for $\gamma$-ray photons due to absorption by the BLR as a function of $\gamma$-ray energy (color scale) and emission region distance from the black hole using \cite{be16}.}
\label{fig:abs}
\end{figure} 
\begin{table}
\begin{minipage}{0.32\textwidth}
\begin{tabular}{ccc} 
E [GeV] &  $\tau$ & $\log{(r\,[cm])}$ \\ 
\hline
151	& 2.88	& $17.64$ \\
185	& 2.61	& $17.76$ \\
227	& 2.77	& $17.74$ \\
277	& 2.77	& $17.74$ \\
339	& 2.78	& $17.76$ \\
450	& 2.58	& $17.78$ \\
\end{tabular}	
\end{minipage}
\begin{minipage}{0.32\textwidth}
\begin{tabular}{ccc} 
E [GeV] &  $\tau$ & $\log{(r\,[cm])}$ \\ 
\hline
167	& 2.61	& $17.74$ \\
245	& 2.71	& $17.78$ \\
471	& 3.28	& $17.68$ \\
\end{tabular}	
\end{minipage}
\begin{minipage}{0.32\textwidth}
\begin{tabular}{ccc} 
E [GeV] &  $\tau$ & $\log{(r\,[cm])}$ \\ 
\hline
168	& 2.58	& $17.74$ \\
246	& 2.84	& $17.75$ \\
369	& 3.07	& $17.75$ \\
550	& 2.35	& $17.82$ \\
\end{tabular}	
\end{minipage}
\caption{Estimates of the distance of the emission region for the three time frames shown in Fig. \ref{fig:abs} left using the expected absorption values shown in Fig. \ref{fig:abs} right. In all tables the energy, the estimated optical depth $\tau$, and the estimated distance from the black hole is shown. \textit{Left:} 2015/05, \textit{middle:} MJD 57161, \textit{right:} MJD 57165.}
\label{tab:tau}
\end{table}
The $\gamma$-ray spectra measured by \textit{Fermi}-LAT and H.E.S.S. during three different integration intervals are shown in Fig. \ref{fig:abs} left. The time frames are MJD 57155.5 - 57170.5 (``2015/05'', black), MJD 57160.4 - 57161.4 (red), and MJD 57164.4 - 57165.4 (blue). All spectral points shown have been corrected for EBL absorption using the model of \cite{fea08}. The \textit{Fermi} spectra are compatible with a log-parabola in all periods. The extrapolations to the VHE range are softer than the EBL-deabsorbed H.E.S.S. spectra (c.f. the dashed black line as an example for the total period). This could be interpreted as an intrinsic absorption feature that has a maximum effect at a few $100\,$GeV. Assuming that this is due to the strong BLR (with a soft photon energy of a few eV), a rough estimate of the location of the VHE emission region can be derived.

The solid lines in Fig. \ref{fig:abs} left show a power-law extrapolation of the \textit{Fermi}-LAT spectrum to VHE. This is a conservative estimate of the intrinsic VHE spectrum. Using then the equation
\begin{eqnarray}
 F_{H.E.S.S.} = F_{Fermi}e^{-\tau} \label{eq:tau},
\end{eqnarray}
where $F_{H.E.S.S.}$ is the measured, EBL-corrected flux and $F_{Fermi}$ is the extrapolated flux in the VHE range, the optical depth $\tau$ can be estimated. Since the parameters of the BLR of \pks are well known, one can determine the theoretical optical depth as a function of distance $r$ from the black hole, as can be seen in Fig. \ref{fig:abs} right using the approach of \cite{be16}. Comparing the derived $\tau$ from Eq. (\ref{eq:tau}) with the theoretical one gives the minimum distance (since the estimate is conservative) of the emission region from the black hole. The results are summarized for the three time frames in Tab. \ref{tab:tau}.

The results give a consistent picture that the VHE emission region is located close to the inner edge of the BLR ($\log{(r_{blr,in}/\mathrm{cm})} = 17.83$). This, however, assumes that the intrinsic spectrum is a power-law, which can be extrapolated from the \textit{Fermi} energy range. If the intrinsic spectrum is much softer than this, the VHE emission region could very well lie beyond the BLR.

\section{DISCUSSION \& CONCLUSIONS}
An exceptional high state of the FSRQ \pks in optical and a moderate high state in the $\gamma$-ray regime led to Target-of-Opportunity observations with the H.E.S.S. array in May 2015. The source was clearly active in the VHE band, and for the first time night-by-night variability has been detected in that energy domain.

The multiwavelength lightcurves are very complex, and no clear correlation patterns are obvious. A much deeper analysis than possible for these proceedings along with a longer-duration data set is necessary to quantify this statement, since usually at least some correlated activity is expected in blazars. Interestingly, this behavior has been observed for \pks on a number of occasions, which led other authors to believe that the inverse Compton hump requires at least two emission mechanisms, if not multiple emission regions. This interpretation is backed up by the data presented here: The strong activity at VHE, coupled with only moderate variability but a hard spectrum at HE, points towards at least two components responsible for the $\gamma$-ray spectrum. The low activity at X-rays also indicates that the spectral component responsible for the X-rays and low-energy $\gamma$-rays is less active than the VHE-emitting component. The strong activity in the optical synchrotron spectrum shows that this part of the synchrotron hump could be related to the electrons producing the VHE component.  Whether the two $\gamma$-ray components are different spectral components by the same electron population (one-zone scenario) or require more electron populations (multi-zone model) cannot be explored without in-depth modeling, which is beyond the scope of these proceedings.

Using the observed VHE flux doubling time of $18\,$hrs, the size of the emission region can be constrained to $R<5.8\times 10^{16} (\delta/40)\,$cm, where $\delta$ is the Doppler factor. Following the standard assumption that the emission region fills the widths of the conical jet, the location can be constrained to $r<0.8 (\Gamma_b/40)^2 (\delta/\Gamma_b)(\Gamma_b\Theta)^{-1}f_{\gamma}^{-1}\,$pc, where $\Gamma_b$ is the Lorentz factor, $\Theta$ the jet opening angle, and $f_{\gamma}$ a scale factor of order unity \cite{aea16}. The absorption study places the emission region within the BLR. Given that this is a conservative lower limit, the combination with the flux doubling time scale estimate places the emission region around the outer edge of the BLR.


All in all, the complex multiwavelength patterns of \pks presented here reveal again the fascinating nature of this FSRQ. It has become even more interesting with the detection of night-by-night variability at VHE $\gamma$-rays by the H.E.S.S. experiment.

\section{ACKNOWLEDGMENTS}
The support of the Namibian authorities and of the University of Namibia in facilitating the construction and operation of H.E.S.S. is gratefully acknowledged, as is the support by the German Ministry for Education and Research (BMBF), the Max Planck Society, the German Research Foundation (DFG), the French Ministry for Research, the CNRS-IN2P3 and the Astroparticle Interdisciplinary Programme of the CNRS, the U.K. Science and Technology Facilities Council (STFC), the IPNP of the Charles University, the Czech Science Foundation, the Polish Ministry of Science and Higher Education, the South African Department of Science and Technology and National Research Foundation, the University of Namibia, the Innsbruck University, the Austrian Science Fund (FWF), and the Austrian Federal Ministry for Science, Research and Economy, and by the University of Adelaide and the Australian Research Council. We appreciate the excellent work of the technical support staff in Berlin, Durham, Hamburg, Heidelberg, Palaiseau, Paris, Saclay, and in Namibia in the construction and operation of the equipment. This work benefited from services provided by the H.E.S.S. Virtual Organisation, supported by the national resource providers of the EGI Federation.
N.C. acknowledges support from Alexander von Humboldt foundation.


\end{document}